\documentclass[12pt,a4]{article}

\usepackage{amsfonts}
\usepackage[margin=2.8cm,nohead]{geometry}

\usepackage[normalem]{ulem}

\newcommand{\hf}{F}

\newcommand{\field}{{\mathbb F}}
\newcommand{\comment}[1]{}

\newcommand{\braket}[2]{{\langle {#1}\!\mid\!{#2} \rangle}}
\newcommand{\Hilbert}{{\cal H}}

\newtheorem{theorem}{Theorem}

\newtheorem{definition}{Definition}
\newtheorem{property}{Property}
\newtheorem{remark}{Remark}

\newcommand{\Endproof}{\hfill$\Box$ \\}

\newcommand{\mat}{\left( \!\! \begin{array}{cc}}
\newcommand{\rix}{\end{array} \!\! \right)}

\usepackage{color}
\def\bR{\begin{color}{red}}
\def\bB{\begin{color}{blue}}
\def\bM{\begin{color}{magenta}}
\def\bC{\begin{color}{cyan}}
\def\bW{\begin{color}{white}}
\def\bBl{\begin{color}{black}}
\def\bG{\begin{color}{green}}
\def\bY{\begin{color}{yellow}}
\def\ec{\end{color}\ }

\usepackage{xy}
\xyoption{matrix}
\xyoption{frame}
\xyoption{arrow}
\xyoption{arc}

\usepackage{ifpdf}
\ifpdf
\else
\PackageWarningNoLine{Qcircuit}{Qcircuit is loading in Postscript mode.  The Xy-pic options ps and dvips will be loaded.  If you wish to use other Postscript drivers for Xy-pic, you must modify the code in Qcircuit.tex}
\xyoption{ps}
\xyoption{dvips}
\fi

\entrymodifiers={!C\entrybox}

\newcommand{\ket}[1]{{\left\vert{#1}\right\rangle}}

\title{Quantum Hashing via Classical $\epsilon$-universal Hashing Constructions}

\author{Farid Ablayev\thanks{Kazan Federal Univesity} \and Marat Ablayev\thanks{Kazan Federal University}}

\date{}

\begin{document}

\maketitle

\begin{abstract}

We define the concept of a quantum  hash generator and offer a design, which allows one to build a large number  of different quantum hash functions. The construction is based on  composition of a classical $\epsilon$-universal hash family and a given family of functions -- quantum hash generators.

The relationship between $\epsilon$-universal hash families and error-cor\-rec\-ting codes give possibilities to build a large amount of different quantum hash functions. In particular, we present  quantum hash function  based on Reed-Solomon code, and we proved, that this construction is optimal in the sense of number of qubits needed.

Using the relationship between $\epsilon$-universal hash families and Frei\-valds' fingerprinting schemas we present explicit quantum hash function and prove that this construction is optimal with respect to the number of qubits needed for the construction.

\end{abstract}




\section{Introduction}

Quantum computing is inherently a very mathematical subject, and the discussions of how quantum computers can be more efficient than classical computers in breaking encryption algorithms  started since Shor invented his famous  quantum  algorithm. The answer of the cryptography community is ``Post-quantum cryptography'', which  refers to research on problems  (usually public-key cryptosystems) that are no more efficiently breakable using quantum computers  than by classical computer architectures. Currently post-quantum cryptography includes several approaches, in particular,  hash-based signature schemes  such as Lamport signatures and Merkle signature schemes.

 Hashing itself is an important basic concept for the organization  transformation and reliable transmission of information. The concept known as ``universal hashing`` was invented  by Carter and Wegman \cite{cw1979} in 1979. In 1994 a relationship  was discovered  between $\epsilon$-universal hash families and error-correcting codes \cite{bjks1994}.  In \cite{avi1995} Wigderson characterizes universal hashing as being a tool which ``should belong to the fundamental bag of tricks of every computer scientist''.

 Gottesman and Chuang proposed a quantum digital system \cite{gc2001}, based on quantum mechanics. Their results are based on quantum a fingerprinting technique and  add  ``quantum direction'' for post-quantum cryptography. Quantum fingerprints have been introduced by Buhrman, Cleve, Watrous and de Wolf in \cite{bcww01}. Gavinsky and Ito \cite{gi2013} viewed quantum fingerprints as cryptographic primitives.

In \cite{av2013,av2014} we considered  quantum fingerprinting as a construction for binary hash functions and  introduced a non-binary  hash function. The  quantum hashing proposed a suitable one-way function for quantum digital signature protocol from \cite{gc2001}.  For more introductory information we refer to \cite{av2013}.

 In this paper, we define the concept of a quantum  hash generator and offer a design, which allows one to build  different quantum hash functions.  The construction is based on the  composition of classical $\epsilon$-universal hash family with a given family of functions -- quantum hash generator.

The  construction proposed combines the properties of robust  presentation   of information by classical error-correcting codes together  with the possibility  of highly compressed presentation of information by quantum systems.

 The relationship between $\epsilon$-universal hash families and error-correcting codes give possibilities to build a large amount of different quantum hash functions. In particular, we present  quantum hash function  based on Reed-Solomon code, and we proved, that this construction is optimal in the sense of number of qubits needed.

Using the relationship between $\epsilon$-universal hash families and Freivalds' fingerprinting schemas we present an explicit quantum hash function and prove that this construction is optimal with respect to of number of qubits needed for the  construction.



\subsection{Definitions and Notations}
We begin by recalling  some definitions of classical hash families from \cite{st1996}.
%
Given a domain $\mathbb X$, $|\mathbb X|=K$, and a range $\mathbb Y$, $|\mathbb Y|=M$, (typically with $K\ge M$), a   hash function $f$ is a map
\[ f : {\mathbb X} \to  {\mathbb Y}, \]
that hash {\em long} inputs to {\em short} outputs.

We let $q$ to be a prime power and $\field_q$ be a finite  field of order $q$. Let $\Sigma^k$  be a set of words of length $k$ over  a finite alphabet $\Sigma$.   In the paper we  let ${\mathbb X}=\Sigma^k$, or ${\mathbb X}=\field_q$, or ${\mathbb X}=(\field_q)^k$, and  ${\mathbb Y}=\field_q$.
A hash family is a set $\hf=\{f_1,\dots, f_N\}$  of  hash functions $ f_i : {\mathbb X} \to  {\mathbb Y}$.

  \paragraph{$\epsilon$ universal  hash family.} A hash family $\hf$ is called an $\epsilon$-universal  hash family if for any two distinct elements $w,w'\in {\mathbb X}$, there exist at most $\epsilon N$ functions $f\in \hf$ such that $f(w)=f(w')$. We will use the notation $\epsilon$-U  $(N;K,M)$ as an abbreviation for $\epsilon$-universal hash family.


 Clearly we have, that if function the $f$ is chosen uniformly at random from a given   $\epsilon$-U   $(N;K,M)$ hash family $\hf$, then the probability that any two distinct words collide under $f$ is at most $\epsilon$.

 The case of  $\epsilon=1/N$ is known as universal hashing.

\paragraph { Classical-quantum function.}\label{cqhf}

The notion of a quantum function was considered  in \cite{mo2010}. In this paper we use the following variant of a quantum function.
 First recall  that mathematically a qubit $\ket{\psi}$ is described as $\ket{\psi}=\alpha\ket{0}+\beta\ket{1}$, where $\alpha$ and $\beta$ are complex numbers, satisfying $|\alpha|^2+|\beta|^2=1$. So, a qubit may be presented  as a unit vector in the two-dimensional Hilbert complex space ${\cal H}^2$.
Let $s\ge 1$. Let $({\cal H}^2)^{\otimes s}$  be the $2^s$-dimensional  Hilbert space, describing the states of $s$ qubits, i.e. $({\cal H}^2)^{\otimes s}$ is made up of $s$ copies of a single qubit space ${\cal H}^2$
 \[({\cal H}^2)^{\otimes s}={\cal H}^2\otimes\dots \otimes{\cal H}^2 = {\cal H}^{2^s}. \]


For $K=|\mathbb X|$ and integer $s\ge 1$ we define a $(K;s)$ classical-quantum  function to be  a
map of the  elements $w\in{\mathbb X}$ to quantum states $\ket{\psi(w)}\in ({\cal H}^2)^{\otimes s}$
\begin{equation}\label{cqf}
\psi : \mathbb X \to ({\cal H}^2)^{\otimes s}.
\end{equation}
We will also use the notation $\psi : w\mapsto \ket{\psi(w)}$ for $\psi$.

%

%

\section{Quantum hashing}

%
%
%
%
%
%
%

What we need to define for quantum hashing and what is
implicitly assumed in various papers (see for example
\cite{av2013} for more information)
is a collision resistance property. However, there is still no such
notion as \emph{quantum collision}. The reason why we need to define
it is the observation that in quantum hashing there might be no
collisions in the classical sense: since quantum hashes are quantum
states they can store an arbitrary amount of data and can be different
for different messages. But the procedure of comparing those quantum
states implies measurement, which can lead to collision-type errors.

So, a \emph{quantum collision} is a situation when a procedure that tests the equality of quantum hashes and outputs ``true'', while hashes are different. This procedure can be a well-known SWAP-test (see for example
\cite{av2013} for more information and citations)  or
something that is adapted for specific hashing function. Anyway, it deals with the notion of distinguishability  of quantum states. Since non-orthogonal quantum states cannot be perfectly distinguished, we require them to be ``nearly orthogonal''.


%

\begin{itemize}
\item For $\delta\in (0,1/2)$ we call a function
\[ \psi : \mathbb X \to ({\cal H}^2)^{\otimes s}\]
$\delta$-resistant, if for any pair $w,w'$ of different elements,
$$
\left|\braket{\psi(w)}{\psi(w')}\right| \le \delta.  \quad
$$
%

\end{itemize}
\begin{theorem}\label{lb} Let
$ \psi : \mathbb X \to ({\cal H}^2)^{\otimes s}$
be a  $\delta$-resistant function. Then
\[s \ge   \log\log {|\mathbb X|} - \log\log\left(1+\sqrt{2/(1-\delta)}\right) -1.\]

\end{theorem}
{\em Proof.} First we observe, that from the definition $||\ket{\psi}||=\sqrt{\braket{\psi}{\psi}}$ of the norm it follows that
 \[
 ||\ket{\psi}-\ket{\psi'}||^2 =||\ket{\psi}||^2 +||\ket{\psi'}||^2 - 2\braket{\psi}{\psi'}.
 \]
Hence for an arbitrary pair $w,w'$ of different elements from $\mathbb X$ we have that
\[ ||\ket{\psi(w)}-\ket{\psi(w')}|| \ge \sqrt{2(1-\delta)}. \]
We let $\Delta = \sqrt{2(1-\delta)}$. For short we let $({\cal H}^2)^{\otimes s} =V$ in this proof.
Consider a set $\Phi=\{ \ket{\psi(w)} : w\in\mathbb X\}$.  If we draw  spheres of  radius $\Delta/2$ with
centres $\ket{\psi}\in \Phi$ then  spheres do not pairwise intersect. All these $K$ spheres are in a large sphere of radius
$1+\Delta/2$. The volume of a sphere
of  radius $r$ in $V$  is $cr^{2^{s+1}}$ for the complex space $V$. The constant $c$ depends on the metric of $V$. From this we have, that the number $K$ is bonded by the number of ``small spheres'' in the ``large sphere''
\[ K\le \frac{c(1+\Delta/2)^{2^{s+1}}}{c(\Delta/2)^{2^{s+1}}}. \]
Hence
\[s \ge   \log\log K - \log\log\left(1+\sqrt{2/(1-\delta)}\right) -1.\]
\Endproof

The notion of  $\delta$-resistance naturally leads to the following  notion of quantum hash function.

\begin{definition}[Quantum hash function]\label{QHF}
Let  $K,s$ be  positive integers and $K=|{\mathbb X}|$. We call a map
%
%
 \[ \psi : {\mathbb X} \to ({\cal H}^2)^{\otimes s} \]
an $\delta$-resistant  $(K;s)$ {quantum hash function} if
 $\psi$ is a $\delta$-resistant function.

We  use the notation $\delta$-R $(K;s)$ as an abbreviation for $\delta$-resistant $(K;s)$ quantum hash functions.

\end{definition}
\section{Generator for Quantum Hash Functions}
In this section  we present  two constructions of quantum hash functions and define  notion of quantum hash function generator, which  generalizes these constructions.


\subsection{Binary quantum hashing.}\label{qf} One of the first explicit quantum hash functions was defined in \cite{bcww01}.   Originally the authors invented a construction called   ``quantum fingerprinting'' for testing the equality of two words for a quantum communication model. The  cryptography aspects of quantum fingerprinting are presented in \cite{gi2013}. The quantum fingerprinting technique  is based on binary error-correcting codes. Later this construction was adopted for cryptographic purposes. Here we present the quantum fingerprinting construction from the quantum hashing point of view.

 An $(n,k,d)$  {\em error-correcting code}  is a map
\[ C : \Sigma^k \to \Sigma^n \]
such that, for any two distinct words $w,w'\in\Sigma^k$,  the Hamming distance   between code words $C(w)$ and $C(w')$ is at least $d$. The code is binary if $\Sigma =\{0,1\}$.

The  construction of a quantum hash function based on quantum fingerprinting  in as follows.
\begin{itemize}
\item Let  $c > 1$ and $\delta < 1$. Let $k$ be a positive integer and $n > k$. Let  $E : \{0, 1\}^k\to \{0, 1\}^n$ be an $(n,k,d)$ binary error-correcting code with  Hamming distance $d\ge(1-\delta)n$.
\item Define a family  of functions $F_E=\{E_1, \dots , E_n\}$, where  $E_i :\{0,1\}^k\to \field_2$ is defined by the rule: $E_i(w)$ is the $i$-th bit of the code word $E(w)$.
\item Let $s=\log{n}+1$. Define the  classical-quantum  function $ \psi_{F_E} :\{0,1\}^k\to({\cal H}^2)^{\otimes s}$, determined  by a word $w$  as
    \[ \psi_{F_E}(w)= \frac{1}{\sqrt{n}}\sum_{i=1}^n\ket{i} \ket{E_i(w)}= \frac{1}{\sqrt{n}}\sum_{i=1}^n\ket{i}\left(\cos\frac{\pi E_i(w)}{2}\ket{0} +\sin\frac{\pi E_i(w)}{2}\ket{1}\right),
     \]
%
\end{itemize}
For $s=\log{n} +1$, the function $\psi_{F_E}$ is an $\delta$-R $(2^k;s)$ quantum hash function,
 that is,  for two different words $w,w'$ we have
\[ |\braket{\psi_{F_E}(w)}{\psi_{F_E}(w')}| \le \delta n/n = \delta. \]

Observe, that the authors in \cite{bcww01} propose, for the first choice of such binary codes,   Justesen
codes with $n=ck$, which give $\delta < 9/10 + 1/(15c)$ for any chosen $c > 2$. Next we observe, that the above construction of a quantum hash function needs $\log{n}+1$ qubits for the fixed $\delta \approx 9/10 + 1/(15c)$. This  number of qubits  is good enough in the sense of the lower bound of Theorem \ref{lb}.

A non-binary quantum hash function is presented in \cite{av2013} and is based on the construction from \cite{av09}.

\subsection{Non-binary quantum hashing.}\label{av-construction}
We present the non-binary quantum hash function from \cite{av2013} in
the following form. For a field $\field_q$, let $B=\{b_1,\dots, b_T\}\subseteq \field_q$. For every $b_j\in
B$ and $w\in\field_q$, define a function $h_j :\field_q \to \field_q$ by the rule
  \[   h_j(w)=b_jw\pmod{q}.\]
  Let $H=\{h_1, \dots h_T\}$ and $t=\log{T}$.
%
%
We define the classical-quantum function
\[ \psi_{H} : \field_q \to ({\cal H}^2)^{\otimes (t+1) } \]
 by the rule
%
\[ \ket{\psi_{H}(w)} =
\frac{1}{\sqrt{T}}\sum\limits_{j=1}^{T}\ket{j}
\left(\cos\frac{2\pi h_j(w)}{q}\ket{0} + \sin\frac{2\pi h_j(w)}{q}\ket{1}\right).
\]
%
 The following is proved in \cite{av2013}.
 \begin{theorem}\label{av}
 Let $q$ be a prime power and $\field_q$ be a field. Then, for arbitrary $\delta >0$, there exists a set $B=\{b_1,\dots, b_T\}\subseteq \field_q$  (and, therefore, a corresponding family $H=\{h_1,\dots,h_T\}$ of functions) with $T=\lceil(2/\delta^2)\ln(2q)\rceil$, such that the quantum function $\psi_{H}$  is a $\delta$-R $(q; t+1)$ quantum hash function.
\end{theorem}

In the rest of the paper we  use the notation $H_{\delta, q}$ to denote this family of functions from Theorem \ref{av} and the notation $\psi_{H_{\delta,q}}$ to denote the corresponding quantum function.

 Observe, that the above  construction of the quantum hash function $\psi_{H_{\delta,q}}$ needs
$t+1
\le \log{\log{2q}}+2\log{1/\delta}+3$ qubits. This  number of qubits  is good enough in the sense of the lower bound of Theorem \ref{lb}.

Numerical results on $\psi_{H_{\delta,q}}$ are presented in \cite{av2013}.

\subsection{Quantum hash generator}
The above two constructions of quantum hash functions are using certain controlled rotations of target qubits.
These transformations are generated by the corresponding
discrete functions from a specific family of functions ($F_E$ and $H_{\delta, q}$ respectively).

These constructions lead to the following definition.
\begin{definition}[Quantum hash generator] \label{q-generator}
Let  $K=|\mathbb X|$ and let ${G}=\{g_1,\dots, g_D\}$ be a family of
functions $g_j :{\mathbb X}\to \field_q$.  Let
$\ell\ge 1$ be an integer. For $g\in {G}$ let   $\psi_{g}$
be a  classical-quantum function $ \psi_{g} :{\mathbb X}\to (\Hilbert^2)^{\otimes\ell}$ determined by the rule
\begin{equation}\label{qg-n}
 \psi_{g} : w \mapsto \ket{\psi_{g}(w)}=\sum_{i=1}^{2^\ell}\alpha_i(g(w))\ket{i},
 \end{equation}
 where the amplitudes $\alpha_i(g(w))$, $i\in\{1,\dots, 2^\ell\}$,  of the state $\ket{\psi_g(w)}$ are  determined by $g(w)$.

 Let $d=\log{D}$. We define a classical-quantum function
$\psi_{G} :{\mathbb X}\to (\Hilbert^2)^{\otimes(d+\ell)} $
by the rule
\begin{equation}\label{qhg}
\psi_{G} : w\mapsto
\ket{\psi_{G} (w)} = \frac{1}{\sqrt{D}}\sum\limits_{j=1}^{D}\ket{j}
\ket{\psi_{g_j}(w)}.
\end{equation}
We say that the family $G$  generates the  $\delta$-R $(K; d+\ell)$ quantum
hash function $\psi_{G}$ and we call $G$ a    $\delta$-R $(K; d+\ell)$
quantum hash generator,  if $\psi_{G}$ is a $\delta$-R $(K; d+\ell)$
quantum hash function.

\end{definition}


According to Definition \ref{q-generator} the family $F_E=\{E_1, \dots, E_n\}$ from Section \ref{qf} is a $\delta$-R $(2^k;\log{n} +1)$ quantum hash generator and the family $H_{\delta,q}$ from Section~\ref{av-construction} is $\delta$-R $(q; t+1)$ quantum hash generator.

%
%
%
%



\section{Quantum Hashing via Classical $\epsilon$-Universal Hashing Constructions}

In this section we present a construction of a  quantum hash generator
based on the composition of an $\epsilon$-universal hash family with a
given  quantum hash generator. We begin with the definitions and notation that we use in the rest of the paper.

Let $K=|\mathbb X|$, $M=|\mathbb Y|$.  Let $ \hf=\{f_1,\dots , f_N \}$ be a family of functions,  where
  \[ f_i : {\mathbb X} \to \mathbb Y.\]

Let $q$  be a prime power and $\field_q$ be a field.  Let $H=\{h_1,\dots , h_T\}$ be a family of functions, where
 \[h_j:\mathbb Y\to \field_q. \]
  For $f\in \hf$ and $h\in H_B$, define  composition $g=f\circ h$,
\[ g :{\mathbb X}\to \field_q, \]
 by the rule
   \[ g(w)=(f\circ h)(w)=h(f(w)). \]
   Define  composition  $G=\hf\circ H$ of two families $F$ and $H$
as follows.
 \[{G}= \{g_{ij}=f_i\circ h_j : i\in I, j\in J\}, \]
where $I=\{1,\dots, N \}$, $J=\{1,\dots, T\}$.


\begin{theorem}\label{main}
 Let $ \hf=\{f_1,\dots , f_N \}$
be  an  $\epsilon$-U $(N;K,M)$ hash family. Let $\ell\ge 1$. Let  $H=\{h_1,\dots h_T\}$
be  a $\delta$-R $(M; \log{T}+\ell)$ quantum hash generator. Let $\log{K} >\log{N}+\log{T} +\ell$.

Then the composition $ {G}=\hf\circ H$ is an $\Delta$-R $(K; s)$ quantum hash generator, where
 \begin{equation}\label{s-th}
 s = \log{N} + \log{T} + \ell
 \end{equation}
 and
\begin{equation}\label{d-th}
\Delta \le \epsilon +\delta.
 \end{equation}


\end{theorem}
{\em Proof.}
The $\delta$-R $(M; \log{T}+\ell)$ quantum hash generator  $H$ generates the  $\delta$-R $(M; \log{T}+\ell)$ quantum hash function
\begin{equation}\label{qh}
 \psi_H :v \mapsto \frac{1}{\sqrt{T}}\sum_{j\in J}\ket{j}\ket{\psi_{h_j}(v)}.
 \end{equation}
For $s=\log{N}+\log{T}+\ell$, using the family $G$,  define the map
\[ \psi_G : \mathbb X \to (\Hilbert^2)^{\otimes s} \]
 by the rule
\begin{equation}\label{qg}
 \ket{\psi_{G} (w)} =\frac{1}{\sqrt{N}}\sum\limits_{i\in I}\ket{i}\otimes \ket{\psi_{H}(f_i(w))}.
 \end{equation}

We show the  $\Delta$ resistance of $\psi_G$.

Consider a pair $w$,$w'$ of different elements from $\mathbb X$ and their inner  product $\braket{\psi_{G}(w)}{\psi_{G}(w')}$. Using the linearity of the inner product we have that
\[ \braket{\psi_{G}(w)}{\psi_{G}(w')}=\frac{1}{N}\sum_{i\in I}\braket{\psi_{H}(f_i(w))}{\psi_{H}(f_i(w'))}.\]
We define two sets of indexes $I_{bad}$  and $I_{good}$:
\[
I_{bad}=\{i\in I : f_i(w)=f_i(w')\}, \quad
I_{good}=\{i\in I : f_i(w)\not=f_i(w')\}.
 \]
  Then we have
\begin{eqnarray}
| \braket{\psi_{G}(w)}{\psi_{G}(w')}| &\le & \frac{1}{N}\sum_{i\in I_{bad}} |\braket{\psi_{H}(f_i(w))}{\psi_{H}(f_i(w'))}| \nonumber \\
 &+ & \frac{1}{N}\sum_{i\in I_{good}} |\braket{\psi_{H}(f_i(w))}{\psi_{H}(f_i(w'))}|. \label{in-p}
 \end{eqnarray}
The hash family $\hf$  is $\epsilon$-universal, hence
\[ |I_{bad}|\le\epsilon N.\]
The quantum function $\psi_H :\mathbb Y\to (\Hilbert^2)^{\log{T}+\ell}$  is $\delta$-resistant, hence for an  arbitrary pair $v$, $v'$ of different elements from $\mathbb Y$ one has

\[|\braket{\psi_{H}(v)}{\psi_{H}(v')}|\le \delta.\]
Finally from (\ref{in-p}) and the above two inequalities we have that
\[ | \braket{\psi_{G}(w)}{\psi_{G}(w')}|\le \epsilon +\frac{|I_{good}|}{N}\delta
 \le \epsilon +\delta.\]
The last inequality  proves $\Delta$-resistance of $\psi_{G}(w)$ (say for $\Delta= \epsilon +\delta(I_{good}|)/N $)   and proves  the inequality (\ref{d-th}).

%

To finish the proof of the theorem it remains to show that the function $\psi_G$
can be presented in the form displayed in (\ref{qhg}). From (\ref{qh}) and (\ref{qg}) we have that
\[\ket{\psi_G(w)}= \frac{1}{\sqrt{N}}\sum\limits_{i\in I}\ket{i}\otimes \left( \frac{1}{\sqrt{T}}\sum\limits_{j\in J}\ket{j}
\ket{\psi_{h_{j}}(f_i(w))}\right). \]
Using the notation from (\ref{qg-n})  the above expression can be presented in the following form (\ref{qhg}).
\[
\ket{\psi_{G} (w)} = \frac{1}{\sqrt{NT}}\sum\limits_{i\in I, j\in J}\ket{ij}
\ket{\psi_{g_{ij}}(w)},
\]
here $\ket{ij}$ denotes a basis quantum state, where $ij$ is treated as a concatenation of the binary representations of $i$ and $j$.
\Endproof

\section{Explicit Constructions of Quantum Hash Functions Based on Classical Universal Hashing}

The following statement is a corollary of Theorem \ref{main} and   a basis for  explicit constructions of quantum hash functions in this section.  Let $q$  be a prime power and $\field_q$ be a field. Let $\delta\in(0,1)$. Let $H_{\delta,q}$ be the family of functions from Theorem~\ref{av}. Let $|\mathbb X|=K$.

\begin{theorem}\label{clhash-qhash}
Let $\hf=\{f_1,\dots, f_N\}$ be an  $\epsilon$-U $(N;K,q)$ hash
family, where $f_i : \mathbb X\to \field_q$. Then for arbitrary $\delta>0$, family $G=\hf\circ H_{\delta,q}$ is a $\Delta$-R $(K;s)$ quantum hash generator, where

\[ s\le \log N + \log\log q +2\log 1/\delta + 3 \]
and
\[ \Delta \le \epsilon +\delta. \]

\end{theorem}
{\em Proof.} We take the family $H_{\delta,q}=\{h_1,\dots, h_T\}$, where $h_i:\field_q\to\field_q$, $T=\lceil(2/\delta^2)\ln(2q)\rceil$, $\ell=1$, and  $s=\log{T}+1 \le \log n + \log\log q +2\log 1/\delta + 3$. $H_{\delta,q}$ is  $\delta$-R $(q; s)$ quantum hash generator. According to Theorem \ref{main} the composition $G=\hf\circ H_{\delta,q}$  is a $\Delta$-R $(K;s)$ quantum hash generator with the stated parameters. 
 \Endproof
\subsection{Quantum hashing from universal linear  hash family}

 The next hash family is  folklore and was  displayed in several papers and books. See the paper \cite{st2002} and the book \cite{st2005b} for more information.



\begin{itemize}
\item Let $k$ be a positive integer and let $q$ be a prime power. Let     ${\mathbb X}=(\field_q)^k\backslash\{(0,\dots,0)\}$. For every vector $a\in (\field_q)^k$ define hash function $f_a:{\mathbb X}\to \field_q$ by the rule
    \[ f_a(w)=\sum_{i=1}^ka_iw_i. \]
    Then $\hf_{lin}=\{ f_a : a\in(\field_q)^k\}$ is an $(1/q)$-U $(q^k;(q^k-1);q)$ hash family (universal hash family).
\end{itemize}


%
\begin{theorem}
 Let $k$ be a positive integer, let $q$ be a prime power.
 Then for arbitrary $\delta\in(0,1)$  composition $G=\hf_{lin}\circ H_{\delta, q}$  is a $\Delta$-R $(q^k;s)$ quantum hash generator with $\Delta \le (1/q) +\delta$ and $s\le k\log{q} +\log{\log{q}} + 2\log{1/\delta} +3$.
\end{theorem}
%
{\em Proof.} According to Theorem \ref{clhash-qhash} function $\psi_{G}$ is  $\Delta$-R $(q^k;s)$ quantum hash function with the parameters stated in the theorem.
\Endproof
\begin{remark}
Note, that from Theorem \ref{lb} we have that
\[ s\ge \log{\log{|{\mathbb X}|}} +\log{\log{\left(1+\sqrt{2/(1-\delta)}\right) }}- 1 \ge  \log{k} + \log{\log{q}} - \log{\log{\left(1+\sqrt{2/(1-\delta)}\right) }}- 1.\]
 This lower bound shows that  the   quantum hash function $\psi_{G}$ is not asymptotically optimal in the sense of number of qubits used for the construction.
\end{remark}




\subsection{Quantum hashing based on Freivalds' fingerprinting}
For  a fixed positive constant $k$  let $\mathbb X =\{0,1\}^k$. Let $c>1$ be a positive integer and let $M =c k\ln{k} $. Let $\mathbb Y=\{0,1,\dots, M-1 \}$.

For the $i$-th prime $p_i\in\mathbb Y$ define a function (fingerprint)
\[f_i : \mathbb X \to \mathbb Y\]
 by the rule
\[ f_i(w)= w \pmod{p_i}. \]
   Here we treat a word $w=w_0w_1\dots w_{k-1}$ also as an integer  $w=w_0+w_12+\cdots + w_{k-1}2^{k-1}$.
Consider the set
\[ F_M=\{ f_1, \dots, f_{\pi(M)} \} \]
of fingerprints. Here $\pi(M)$ denotes the number of primes less than  or equal to $M$.
Note that then $\pi(M)\sim M/\ln{M}$ as $M\to\infty$.
 Moreover,
 \[  \frac{M}{\ln{M}}\le \pi(M)\le 1.26\frac{M}{\ln{M}} \qquad \mbox{for} \quad M\ge 17. \]

The following fact is based on a construction, ``Freivalds' fingerprinting method'', due to  Freivalds \cite{f1977}.
\begin{property}\label{freivald}
The set $F_M$ of fingerprints is a $(1/c)$-U $(\pi(M);2^k,M)$ hash family.
\end{property}
{\em Proof (sketch).} For any pair $w$, $w'$ of distinct words from $\{0,1\}^k$ the number $N(w,w')= |\{f_i\in F_M : f_i(w)=f_i(w')\}|$ is bounded from above by $k$.
Thus,  if we pick a prime $p_i$ (uniformly at random) from $\mathbb Y$ then
\[ Pr[f_i(w)=f_i(w')]\le \frac{k}{\pi(M)}\le \frac{k\ln{M}}{M}. \]
Picking $M=ck\ln{k}$ for a constant $c$ gives $ Pr[f_i(w)=f_i(w')]\le \frac{1}{c}+o(1)$.
 \Endproof

Theorem \ref{clhash-qhash} and Property \ref{freivald}  provide  the following statement.

\begin{theorem}
Let $c>1$ be a positive integer and let $M =c k\ln{k} $. Let $q\in \{M,\dots, 2M\}$ be a prime. Then, for arbitrary $\delta>0$, family $G=F_M\circ H_{\delta,q}$ is a $\Delta$-R $(2^k;s)$ quantum hash generator, where
\[ s\le \log{ck}+\log{\log{k}} +\log{\log{q}} +2\log{1/\delta}+3 \]
and
\[ \Delta\le \frac{1}{c} +\delta.\]

\end{theorem}
{\em Proof.} From Theorem \ref{clhash-qhash} we have that
\[ s\le \log{\pi(M)} +\log\log{q} +2\log{1/\delta}+3. \]
From the choice of $c$ above  we have that $M =c k\ln{k} $. Thus
\[ s\le \log{ck}+\log{\log{k}} +\log{\log{q}} +2\log{1/\delta}+3. \]
\Endproof
\begin{remark}
Note that from Theorem \ref{lb} we have
\[ s\ge \log{k} + \log{\log{q}} - \log{\log{\left(1+\sqrt{2/(1-\delta)}\right) }}- 1.\]
 This lower bound shows that  the   quantum hash function $\psi_{\hf_M}$ is good enough in the sense of the number of qubits used for the construction.
\end{remark}


\subsection{Quantum hashing  and error-correcting codes}
Let $q$ be a prime power and let $\field_q$ be a field.
An $(n,k,d,)$   error-correcting code is called  {\em linear}, if $\Sigma=\field_q$, and ${\cal C}=\{C(w) : w\in\field_q^k\}$ is a subspace of $(\field_q)^n$. We will denote such linear code by an  $[n,k,d,]_q$ {code}.

\begin{theorem}\label{code-qhg}
 Let $\cal C$ be an $[n,k,d]_q$ code. Then for arbitrary $\delta\in (0,1)$ there exists a $\Delta$-R $(q^k;s)$ quantum hash generator $G$, where $\Delta=(1-d/n)+\delta$ and $s\le \log n  +\log\log q +2\log{1/\delta} +4$.
\end{theorem}
{\em Proof.} The following fact was observed in \cite{bjks1994,st1996}. Having an $[n,k,d]_q$ code $\cal C$, we can explicitly construct  a $(1-d/n)$-U $(n;q^k;q)$ hash family $F_{\cal C}$.

By Theorem \ref{clhash-qhash} a  composition $G=F_{\cal C}\circ H_{\delta, q}$ is an $\Delta$-R $(q^k;s)$ quantum hash generator, where $\Delta=(1-d/n)+\delta$ and $s\le \log n  +\log\log q +2\log{1/\delta} +4$. \Endproof


\subsubsection{Quantum hash function via Reed-Solomon code}
As an example we present construction of quantum hash function, using Reed-Solomon codes.

 Let $q$ be a prime power, let $k\le n\le q$, let $\field_q$ be a finite field. A {\em  Reed-Solomon} code (for short RS-code) is a linear code
 \[ C_{RS}:(\field_q)^k\to (\field_q)^n \]
  having parameters $[n,k,n-(k-1)]_q$. RS-code defined as follows. Each word $w\in(\field_q)^k$, $w=w_0w_1\dots w_{k-1}$ associated with the polynomial
\[ P_w(x)= \sum_{i=0}^{k-1}w_ix^i. \]
Pick $n$ distinct elements (evaluation points) $A=\{a_1,\dots, a_{n}\}$  of $\field_q$. A common special case is $n=q-1$ with the set of evaluating points being $A=\field_q\backslash\{0\}$. To encode word $w$ we evaluate $P_w(x)$ at all $n$ elements $a\in A$

\[  C_{RS}(w)=(P_w(a_1)\dots P_w(a_{n})). \]
Using Reed-Solomon codes, we obtain the following construction of quantum hash generator.
\begin{theorem}
Let $q$ be a prime power and let $1\le k\le n\le  q$. Then for arbitrary $\delta\in (0,1)$ there is a $\Delta$-R $(q^k;s)$ quantum hash  generator $G_{RS}$, where $\Delta\le \frac{k-1}{n}+\delta$ and  $s\le \log {(q\log q)} +2\log1/\delta +4$.
\end{theorem}
{\em Proof.} Reed-Solomon code $C_{RS}$ is $[n,k,n-(k-1)]_q$ code, where $k\le n\le q$. Then according to Theorem \ref{code-qhg} there is a  family $G_{RS}$, which is an $\Delta$-R $(q^k;s)$ quantum hash generator  with stated parameters. \Endproof

 In particular, if we select  $n\in [ck, c'k]$ for constants $c<c'$, then  $\Delta \le 1/c +\delta$ for $\delta\in(0,1)$ and in according to  Theorem  \ref{lb} we get that

%
\[  \log {(q\log q)} - \log\log\left(1+\sqrt{2/(1-\Delta)}\right) -\log {c'/2}\le s\le \log {(q\log q)} +2\log1/\Delta +4. \]
 Thus, Reed Solomon codes provides good enough parameters for resistance value $\Delta$  and  for a number $s$ of qubits we need to construct quantum hash function $\psi_{RS}$.

 \paragraph{Explicit constructions of $G_{RS}$ and $\psi_{G_{RS}}$.}
 Define $(k-1)/q$-U $(q;\field_q^k;q)$ hash family  $F_{RS}=\{ f_a :  a\in A\}$ based on $C_{RS}$ as follows. For $a\in A$ define $f_a: (\field_q)^k \to \field_q$ by the rule
\[ f_a(w_0\dots w_{k-1})= \sum_{i=0}^{k-1}w_ia^i. \]
Let $H_{\delta, q}= \{h_1,\dots, h_T\}$, where $h_j :\field_q\to\field_q$ and  $T=\lceil(2/\delta^2)\ln{2q}\rceil$.
For $s=\log{n} +\log{T}+1$ composition $G_{RS}=F_{RS}\circ H_{\delta, q}$,  defines  function \[\psi_{G_{RS}} :(\field_q)^k\to (\Hilbert^2)^{\otimes s}\]
 for a word $w\in (\field_q)^k$ by the rule.
 \[
\psi_{G_{RS}}(w)=   \frac{1}{\sqrt{n}}\sum_{i=1}^n\ket{i}\otimes \left(\frac{1}{\sqrt{T}}\sum_{j=1}^{T}\ket{j}\left(\cos\frac{2\pi h_j(f_{a_i}(w))}{q}\ket{0} + \sin\frac{2\pi h_j(f_{a_i}(w))}{q}\ket{1} \right)\right).
 \]



\end{document}